\documentclass{article}
\usepackage{LaThuileFPSpro}
\usepackage{graphicx, url} % Include figure files
\def\Journal#1#2#3#4{{#1} {\bf #2}, #3 (#4)}

\def\NPB{{\em Nucl.\ Phys.\ } B}
\def\PLB{{\em Phys.\ Lett.\ }  B}
\def\PRL{\em Phys.\ Rev.\ Lett.\ }
\def\PRD{{\em Phys.\ Rev.\ } D}
\def\RMP{\em Rev.\ Mod.\ Phys.\ }
\def\ZPC{{\em Z.\ Phys.\ } C}

\def\ra{\rightarrow}

\def\be{\begin{equation}}
\def\ee{\end{equation}}
\def\bea{\begin{eqnarray}}
\def\eea{\end{eqnarray}}

\def\elel{\ell^+\ell^-}

\def\KP{K^+}
\def\piP{\pi^+}
\def\piM{\pi^-}
\def\piZ{\pi^0}

\def\KS{K^0_S}

\def\Kst{K^*}
\def\KstZ{K^{*0}}
\def\KstP{K^{*+}}

\def\Kstll{\Kst\elel}

\def\TeV{{\rm~TeV}}

\def\fbinv{{\rm~fb}^{-1}}

\def\Mbc{M_{\rm bc}}

\def\Cseven{\tilde{C}_7^{\rm eff}}
\def\Cnine{\tilde{C}_9^{\rm eff}}
\def\Cten{\tilde{C}_{10}^{\rm eff}}
\def\Ci{\tilde{C}_i}

\def\PM#1#2{\,^{+#1}_{-#2}}

\def\Journal#1#2#3#4{{#1} {\bf #2}, #3 (#4)} % {journal}{vol}{page}{year}
\def\NPB{Nucl. Phys. B}
\def\PLB{Phys. Lett. B}
\def\PRL{Phys. Rev. Lett.}
\def\PRD{Phys. Rev. D}
\def\ZPC{Z. Phys. C}

\def\RMP{Rev. Mod. Phys.}

\def\l{\ell}

\def\blv{B \rightarrow  \l \, \nu}
\def\bll{B^0 \rightarrow \l^+ \, \l^- }

\def\afb{{{\cal A}_{\rm FB}}}

\def\be{\begin{equation}}
\def\ee{\end{equation}}
\def\ba{\begin{eqnarray}}
\def\ea{\end{eqnarray}}

\begin{document}
\title{ \boldmath
  BEYOND THE STANDARD MODEL AT BELLE: $B \ra K^{*} \ell^{+} \ell^{-} $ 
  AND SEARCH FOR LEPTONIC $B$ DECAYS
  }
\author{
  Stefano Villa\\
  {\em Laboratory for High-Energy Physics,
  Ecole Polytechnique F\'ed\'erale,}\\ 
  {\em CH-1015 Lausanne, Switzerland} 
  }
  
\maketitle 

\baselineskip=11.6pt

\begin{abstract}
We report the first measurement of the forward-backward asymmetry and 
the ratios of Wilson coefficients
$A_9/A_7$ and $A_{10}/A_7$ 
in $B \to \Kstll$, obtained using 386M $B\overline{B}$ pairs 
that were collected at the $\Upsilon(4S)$ 
resonance with the Belle detector at the KEKB asymmetric-energy $e^+ e^-$ collider.
We also summarise the results obtained by Belle in searches for purely leptonic
$B$ decays, with emphasis on their impact on models of physics 
beyond the Standard Model.
\end{abstract}
\newpage
\section{Introduction}
The Belle detector\cite{belle}, 
operating at the KEKB $e^+e^-$ collider\cite{KEKB} 
at a centre-of-mass energy corresponding 
to the mass of the $\Upsilon(4S)$ resonance, 
has accumulated as of May 2006 a data sample
corresponding to more than 500M $B\overline{B}$ pairs.
Such a large sample of $B$ mesons allows the study of
rare $B$ decays, which are among the cleanest probes
of the flavour sector of the Standard Model (SM) available to present
experiments.
In particular, these decays have the potential of revealing the existence
of new particles and couplings not present in the SM, 
but predicted by several models of physics Beyond the SM (BSM).

In the last few years Belle has reported several measurements that 
constrain the parameter space of BSM theories.
In particular, this review will concentrate on the measurement of 
forward-backward asymmetry and of ratios of Wilson coefficients 
in $B \to \Kstll$\cite{bksll-paper}, described in Section~\ref{Sec:bksll}, 
and on the searches for purely leptonic $B$ decays of the type
$\blv$, detailed in Section~\ref{Sec:blv}, and $\bll$ 
(Section ~\ref{Sec:bll}). 
%
%%%%%%%%%%%%%%%%%%%%%%%%%%%%%%%%%%%%%%%%%%%%%%%%%%%%%%%%%%%%%%%%%%%%%%%%
\section{\boldmath Measurement of forward-backward asymmetry and 
ratios of Wilson coefficients in $B \to \Kstll$}\label{Sec:bksll}
%%%%%%%%%%%%%%%%%%%%%%%%%%%%%%%%%%%%%%%%%%%%%%%%%%%%%%%%%%%%%%%%%%%%%%%%
%
Flavor-changing neutral current $b \to s$ processes are forbidden at 
tree level and can only proceed 
via loop diagrams in the SM. 
Loops are sensitive to new physics effects via insertion of 
heavy particles in the internal lines; if new heavy particles can
contribute to the decays, their amplitudes will interfere with the SM amplitudes
and thereby modify the decay rate as well as differential distributions.

Such contributions would therefore change the
Wilson coefficients\cite{BBL} that parametrise the strength of the short
distance interactions. 
The $b \to s \ell^+ \ell^-$ amplitude is described
by the effective Wilson coefficients $\Cseven$, $\Cnine$ and 
$\Cten$, whose terms have been calculated up to 
next-to-next-to-leading order (NNLO) in the SM\cite{WC}. 
A measurement of the forward-backward asymmetry and of the 
differential decay rate 
$g(q^2,\theta)= d^2\Gamma/dq^2d\cos\theta$
as functions of $q^2$ and $\theta$ for $B \to \Kstll$ 
constrains the relative 
signs and magnitudes of these coefficients\cite{C10flip,C7C9C10}. 
Here $q^2$ is the squared invariant mass of  the dilepton system, 
and $\theta$ is the angle between the momenta of the negatively (positively)
charged lepton and the $B$ ($\overline{B}$) meson in the dilepton rest frame.
The forward-backward asymmetry is defined as
\begin{equation}
{\cal A}_{\rm FB}(q^2) ={  \int^1_{-1} {\rm sgn}(\cos\theta)g(q^2,\theta) d\cos\theta \over  \int^1_{-1} g(q^2,\theta) d\cos\theta }.
\end{equation}
%
%%%%%%%
\subsection{Selection of $B \to \Kstll$ events}
%%%%%%%
%
The analysis is based on a $357\fbinv$ data sample containing 386M 
$B\overline{B}$ pairs; the $B^+ \to K^+ \ell^+ \ell^-$ 
mode is also studied, since it 
is expected to have a very small forward-backward 
asymmetry even in the presence of
new physics\cite{NoAFBKll} and can therefore be used as null test for the
$\afb$ measurement.

The following final states are used to reconstruct $B$ candidates:
$\KstZ \elel$, $\KstP \elel$, and $\KP \elel$,
with subdecays
$\KstZ\to\KP\piM$, $\KstP\to\KS\piP$ and $\KP\piZ$,
$\KS\to\piP\piM$, and $\piZ\to\gamma\gamma$.
Hereafter, $\KstZ \elel$ and $\KstP \elel$ are combined and called
$K^* \elel$.
Two variables defined in the centre-of-mass (CM) frame are used
to select $B$
candidates: the beam-energy constrained mass $M_{\rm bc}\equiv\sqrt{E_{\rm beam}^{2} - p_{B}^{2}}$
and the energy difference $\Delta E\equiv E_{B} - E_{\rm beam}$, where 
$p_{B}$ and $E_B$ are the measured CM momentum and energy of the $B$
candidate, and $E_{\rm{beam}}$ is the CM beam energy. 

%
%%%  BACKGROUND  %%%
%
The dominant background consists of $B\overline{B}$ events where both $B$ 
mesons decay semileptonically. 
This background is suppressed using missing energy and $\cos\theta_B^*$, where 
$\theta_B^*$ is the angle between 
the flight direction of the $B$ meson and the beam axis in the CM frame.
The resonant contributions due to charmonia ($J/\psi$ and $\psi(2S)$) 
are rejected using the
dilepton invariant mass.
The signal box is defined as $|M_{\rm bc}-m_{B}|<8$~MeV/$c^{2}$ for
both lepton modes and
$-55\, (-35)\, {\rm MeV} < \Delta E < 35\, {\rm MeV}$ for the electron (muon) mode.
%
%%%  FITTING  %%%
%
An unbinned maximum-likelihood fit to the $M_{\mathrm{bc}}$ distribution 
is performed in order to determine the signal yield.
The fit includes several event categories, i.e. signal, 
three types of cross-feeds and four background sources. 
In the fit, all background fractions except the dilepton background are fixed 
while the signal fraction is allowed to float.

\begin{figure}[htbp]
\begin{center}
%  \vspace{9.0cm}
%  \special{psfile=fig1.eps 
%    hscale=50 vscale=50 angle=0}
  \includegraphics[scale=0.5]{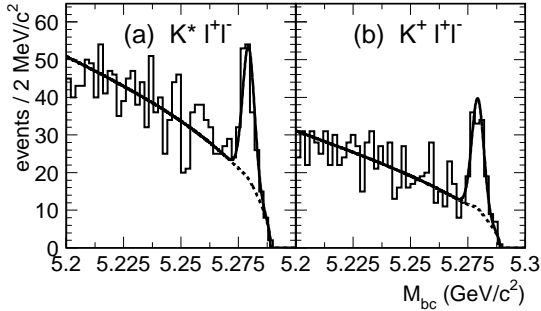}
\end{center}
\caption{\em $M_{\rm bc}$ distributions for the 
(a) $B \to K^{*} \ell^{+} \ell^{-}$ 
and (b) $B\to K^+ \ell^{+} \ell^{-}$ samples. The solid
and dashed curves are the fit results for the total and 
background contributions, respectively.}
\label{fig:mbcfit}
\end{figure}

%
%%%  YIELD  %%%
%
The fit yields $113.6 \pm 13.0$ and $96.0 \pm 12.0$ signal events for the
$\Kstll$ and $K^+\ell^+\ell^-$ modes, respectively.
Data and fit results are shown in Figure~\ref{fig:mbcfit}. 
%
%%%%%%%
\subsection{Extraction of $\afb$ and Wilson coefficients}
%%%%%%%
%
The $B \to K^* \ell^+ \ell^-$ candidates in the signal box are used to measure 
the normalised double differential decay width. 
For the evaluation of the Wilson coefficients, the NNLO Wilson coefficients 
$\Ci$ of Ref.\cite{WC} are used. 
Since the full NNLO calculation only exists for $q^2/m_b^2<0.25$, 
we adopt the so-called partial NNLO calculation\cite{ALGH} for $q^2/m_b^2>0.25$.
The higher order terms in the $\Ci$ are fixed to the SM values while 
the leading terms $A_i$, with the exception of $A_7$, are allowed to float. 
Since the branching fraction measurement of 
$B \to X_s \gamma$ is consistent with the prediction for $|A_7|$
within the SM, $A_7$ is fixed at the SM value, 
$-0.330$, or the sign-flipped value, $+0.330$. 
The fit parameters are therefore $A_9/A_7$ and $A_{10}/A_7$;
the SM predictions for $A_9$ and $A_{10}$ are 4.069 and --4.213, 
respectively\cite{ALGH}.

To extract these ratios, an unbinned maximum likelihood fit is performed 
to the events in the signal box with a probability density function
that includes the normalised double differential decay width.

The $q^2$-integrated asymmetry $\tilde{{\cal A}}_{\rm FB}$ is measured by
determining the yield in forward and backward regions from a fit to the 
$\Mbc$ distribution. 
After correcting for efficiency, the result is:
\begin{eqnarray}
\tilde{{\cal A}}_{\rm FB}(B \to \Kstll) &=& 0.50 \pm 0.15 \pm 0.02, \nonumber \\
\tilde{{\cal A}}_{\rm FB}(B^+ \to K^+ \ell^+ \ell^-) &=& 0.10 \pm 0.14 \pm 0.01,
\end{eqnarray}
where the first uncertainty is statistical and the second is systematic.
A large integrated asymmetry is observed for $\Kstll$ with a significance of $3.4\sigma$.
The result for $K^+ \ell^+ \ell^-$ is consistent with zero as expected. 

\begin{figure}[htbp]
\begin{center}
  \includegraphics[scale=0.65]{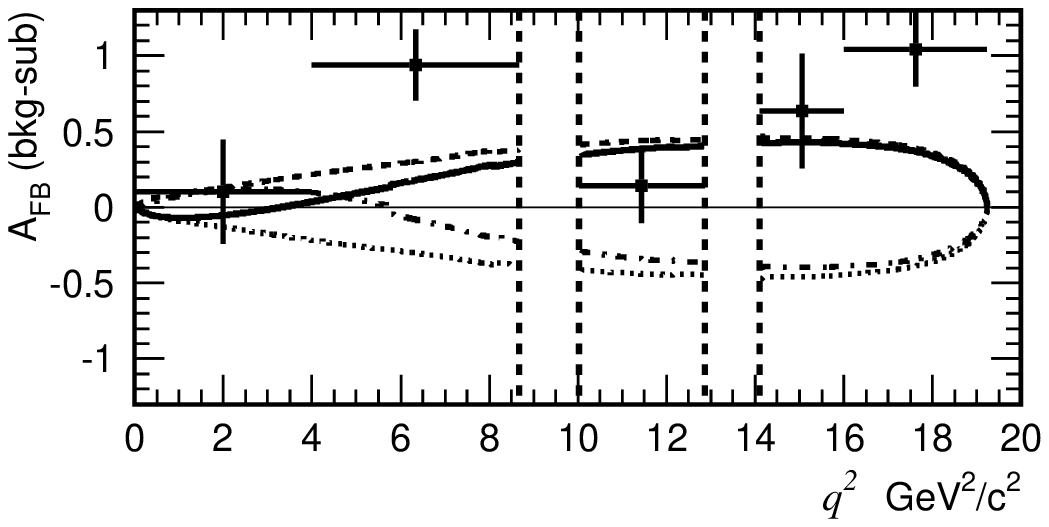}
\end{center}
\caption{\em Fit result for the negative $A_7$ solution (solid) projected 
  onto the background subtracted forward-backward asymmetry, 
  and forward-backward asymmetry curves for several input parameters, 
  including the effects of efficiency: $A_7$ positive 
  case~($A_7=0.330$, $A_9=4.069$, $A_{10}=-4.213$) (dashed), 
  $A_{10}$ positive case ($A_7=-0.280$, $A_9=2.419$, $A_{10}=1.317$) 
  (dot-dashed) and both $A_7$ and $A_{10}$ positive case~($A_7=0.280$, $A_9=2.219$, 
  $A_{10}=3.817$) (dotted). 
  The BSM scenarios shown by the dot-dashed and dotted curves are excluded
  by this measurement.}
\label{fig:afb}
\end{figure}

The fit results of ratios of Wilson coefficients are summarised in Table~\ref{tab:result}.
Figure~\ref{fig:afb} shows the fit results projected onto the 
background-subtracted forward-backward asymmetry distribution in bins of $q^2$. 
\begin{table}[htb]
  \caption{\em
    $A_9/A_7$ and $A_{10}/A_7$ fit results for negative and positive $A_7$ values. 
    The first uncertainty is statistical and the second is systematic.}
 \begin{center}
 \renewcommand{\arraystretch}{1.2}
 \begin{tabular}{ccccc} \hline 
		 & & Negative $A_7$       & & Positive $A_7$ \\ \hline
$A_9/A_7$        &  & $-15.3 \PM{3.4}{4.8}\pm1.1$ &  & $-16.3 \PM{3.7}{5.7}\pm1.4$ \\ 
$A_{10}/A_7$     &  & $10.3 \PM{5.2}{3.5}\pm1.8$  &  & $11.1 \PM{6.0}{3.9}\pm2.4$ \\ \hline 
  \end{tabular}
   \end{center}
\label{tab:result}
\end{table}

\begin{figure}[htb]
\begin{center}
  \includegraphics[scale=0.4]{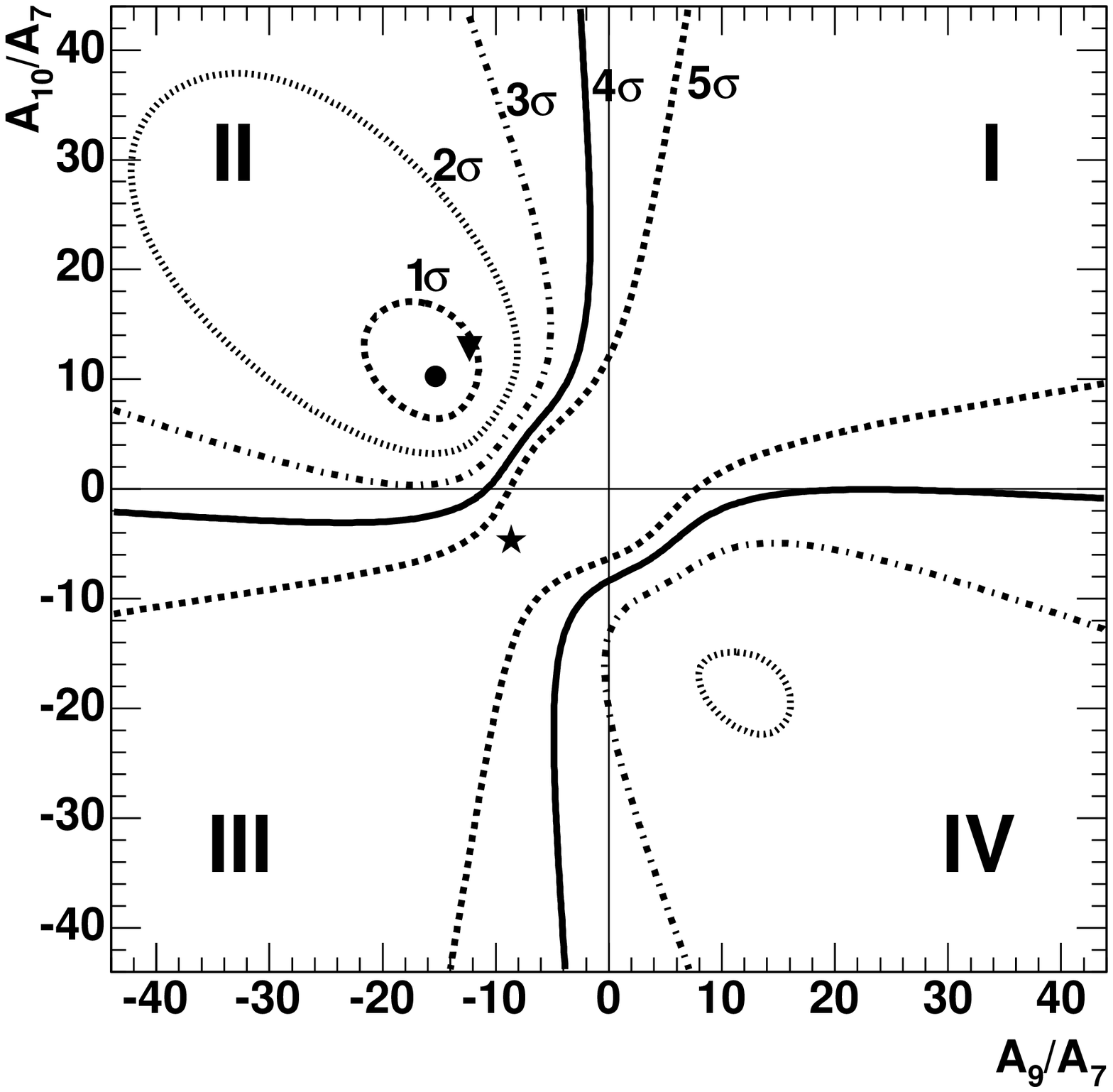}
\end{center}
\caption{\em
  Confidence level contours for negative $A_7$. Curves show 1$\sigma$ to 5$\sigma$ contours. 
  The symbols show the fit (circle), SM (triangle), and $A_{10}$-positive (star) cases.}
\label{fig:cl}
\end{figure}

The fit results are 
consistent with the SM values ${A_9}/{A_7}=-12.3$ and ${A_{10}}/{A_7}=12.8$.
In Figure~\ref{fig:cl}, we show confidence level (CL) contours in the 
($A_9/A_7$, $A_{10}/A_7$) plane 
based on the fit likelihood smeared by the systematic uncertainty, 
which is assumed to have a Gaussian distribution.
We also calculate an interval in $A_9A_{10}/A_7^2$ at the 95\% CL for the 
allowed $A_7$ region,
\begin{eqnarray}
-14.0 \times 10^2 < {A_9}{A_{10}}/{A_7^2} < -26.4.
\end{eqnarray}
This result implies that the sign of ${A_9}{A_{10}}$ must be negative, 
and the solutions in quadrants I and III 
of Figure~\ref{fig:cl} are excluded at 98.2\% CL.
Since solutions in both quadrants II and IV are allowed, we cannot 
determine the sign of $A_7A_{10}$.

%
%%%%%%%%%%%%%%%%%%%%%%%%%%%%%%%%%%%%%%%%%%%%%%%%%%%%%%%%%%%%%%%%%%%%%%%%
\section{\boldmath Search for $\blv$ }\label{Sec:blv}
%%%%%%%%%%%%%%%%%%%%%%%%%%%%%%%%%%%%%%%%%%%%%%%%%%%%%%%%%%%%%%%%%%%%%%%%
%

Leptonic decays of charged $B$ mesons proceed in the SM via the 
$W$-mediated annihilation tree diagram, with a branching fraction given by:
\be \label{eq:B2lnu}
\mathcal{B}(B^+\rightarrow \ell^+\nu_\ell) =
\frac{G_F^2 m_B}{8\pi} m_\ell^2 \left(1-\frac{m_\ell^2}{m_B^2}\right)^2
f_B^2 |V_{ub}|^2 \tau_B , 
\ee
where $\tau_B$ is the $B$ meson lifetime, $f_B$
is the $B$ decay constant and $V_{ub}$ an element of the CKM-matrix.
These modes are thus very interesting because they 
give direct access to the product  $f_B \times V_{ub}$, from which
one can extract a measurement of $f_B$.
The SM expectation for $B^+\rightarrow \tau^+\nu_\tau$ is around
${\cal B}(B^{+}\rightarrow\tau^{+}\nu_{\tau}) 
= 1.0 \times 10^{-4}$; decays to lighter leptons are helicity suppressed, 
by factors of 223 for muons and 10$^7$ for electrons.
Allowing for decay amplitudes BSM, measurements of
these processes give stringent limits on important parameters of such
SM extensions, e.g.\ the mass of the charged Higgs boson and 
$\tan\beta$ (the ratio of vacuum expectation values of the two Higgs 
doublets) in the minimal supersymmetric SM (MSSM), or leptoquark masses in 
Pati-Salam models.

%%%%%%%
\subsection{$B^+\rightarrow \tau^+\nu_\tau$}
\begin{figure}[htb]
\begin{center}
  \includegraphics[scale=0.35]{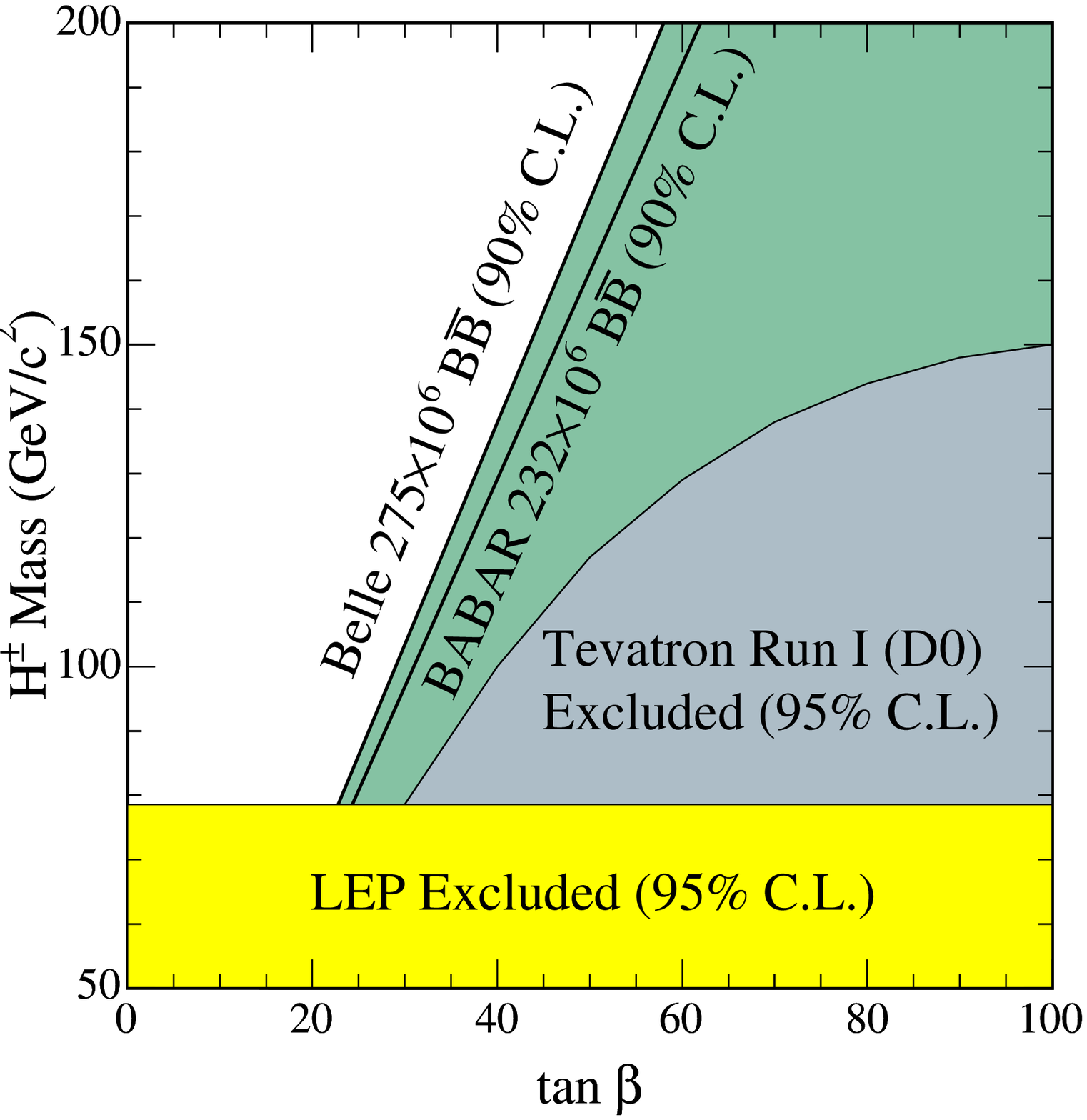}
\end{center}
\caption{\em The $90\%$ C.L. exclusion boundaries in the 
	$[m_{H^{+}}, \tan\beta]$ plane obtained from the observed upper 
	limit on ${\cal B}(B^{+}\rightarrow \tau^{+}\nu)$, compared with 
	other experimental searches.}
\label{fig:btaunu}
\end{figure}

%%%%%%%
The search for $B^+\rightarrow \tau^+\nu_\tau$\footnote{A few weeks after this conference, 
the Belle collaboration has
announced evidence for $B^+\rightarrow \tau^+\nu_\tau$\cite{hep-ex/0604018},
thus providing the first evidence of a purely leptonic $B$ decay and the
first direct determination of the decay constant $f_B$.}
decays is based on the
full reconstruction of one of the $B$ mesons in 
the event, in the modes
$B^+\rightarrow D^{(*)0}h^+$ and $D^{(*)0} D_s^{(*)+}$ 
($h = \pi, K$), selected in a sample of 275M $B\overline{B}$ events\cite{hep-ex/0507034}. 
Decays of the types $\tau\rightarrow \mu(e)\nu\bar{\nu}$, $\pi\nu$, $\pi\pi^0\nu$, 
and $\pi\pi\pi\nu$ are then searched for in the event remainders.
The final event selection is based on the extra energy present in the
electromagnetic calorimeter. 
So-called double-tag events (i.e.\ fully reconstructed $\Upsilon(4S)$ 
decays) are used to validate the simulation of this quantity.
The resulting upper limit on the branching fraction is 
${\cal B}(B^{+}\rightarrow\tau^{+}\nu_{\tau}) 
< 1.8 \times 10^{-4}$ at 90\% CL.
In the two-Higgs doublet model, this limit can be translated\cite{Hou:1992sy} in the 
exclusion of a region in the Higgs mass ($m_{H^{+}}$) versus $\tan\beta$ plane. The
Belle result is shown in Figure~\ref{fig:btaunu} together with
constraints placed by other 
measurements\cite{Bock:2000gk, Abazov:2001md, PRD73-057101-2006}.
%
%%%%%%%
\subsection{$B^+\rightarrow \mu^+\nu_\mu$ and $B^+ \rightarrow e^+\nu_e$}
%%%%%%%
The electron and muon channels are studied in Belle by requiring one 
highly energetic lepton in the event and large missing energy and 
missing momentum corresponding to the undetected neutrino. All
other particles in the event are used to reconstruct the companion
$B$, which has to satisfy requirements on $M_{\rm bc}$ and $\Delta E$. 
The final selection variable used to isolate the signal is the 
lepton momentum in the $B$ rest frame, defined by the momentum
of the recoiling companion $B$.
The latest Belle results are 
${\cal B}(B^{+}\rightarrow\mu^{+}\nu_{\mu}) 
< 2.0 \times 10^{-6}$ at 90\% CL, based on a data sample 
of $140 \fbinv$\cite{b2munu} and
${\cal B}(B^{+}\rightarrow e^{+}\nu_e) 
< 5.4 \times 10^{-6}$ at 90\% CL, based on $60 \fbinv$\cite{b2enu}.
%
%%%%%%%%%%%%%%%%%%%%%%%%%%%%%%%%%%%%%%%%%%%%%%%%%%%%%%%%%%%%%%%%%%%%%%%%
\section{\boldmath Search for $\bll$ }\label{Sec:bll}
%%%%%%%%%%%%%%%%%%%%%%%%%%%%%%%%%%%%%%%%%%%%%%%%%%%%%%%%%%%%%%%%%%%%%%%%
%
The decay of $B^0$ to pairs of charged leptons can proceed in the SM
via box diagrams or via annihilation into $Z$ boson or photon.
Similarly to their $B^\pm$ counterparts, the predicted branching fractions
are helicity suppressed for light leptons, down to about $10^{-10}$ for
muons and  $10^{-15}$ for electrons.
Enhancements of these branching fractions are predicted 
by several BSM theories, such as high-$\tan\beta$ MSSM and some SUSY
scenarios. The lepton flavour violating mode 
$B^0 \rightarrow e^{\pm} \, \mu^{\mp}$ is forbidden in the SM and so 
any signal in this mode would be an undeniable sign of new physics.
Belle has set the following 90\% CL upper limits based on a data 
sample of 78~$\fbinv$\cite{bll}:
${\cal B}(B^0 \rightarrow \mu^{+} \mu^{-}) < 1.6 \times 10^{-7}$,  
${\cal B}(B^0 \rightarrow e^{+} e^{-}) < 1.9 \times 10^{-7}$ and 
${\cal B}(B^0 \rightarrow e^{\pm} \, \mu^{\mp}) < 1.7 \times 10^{-7}$.
The latter result can be interpreted as a lower bound on the Pati-Salam
leptoquark mass\cite{pane-salame}: $m_{\mathrm{LQ}} > 46 \TeV/c^2$ at 90\% CL.
%
%%%%%%%%%%%%%%%%%%%%%%%%%%%%%%%%%%%%%%%%%%%%%%%%%%%%%%%%%%%%%%%%%%%%%%%%

%
\end{document}